\documentclass[a4paper]{jpconf}

\usepackage{graphicx}

\begin{document}
\title{Thouless-Valatin moment of inertia and removal of the spurious mode in the linear response theory}

\author{Markus Kortelainen}

\address{Department of Physics, P.O. Box 35 (YFL), University of Jyvaskyla, FI-40014 Jyvaskyla, Finland\\
         Helsinki Institute of Physics, P.O. Box 64, FI-00014 University of Helsinki, Finland}

\ead{markus.kortelainen@jyu.fi}

\begin{abstract}
Symmetry breaking at the mean-field level leads to an appearance of a symmetry restoring Nambu-Goldstone (NG) mode
in the linear response theory. These modes represent a special kind of collective motion of the system. However, 
they can interfere with the calculated intrinsic physical excitations and, hence, they are often called as spurious modes.
I discuss translational and rotational NG mode and the inertia parameter associated with these modes,
by using the finite amplitude method formalism. I will also discuss how to remove spurious mode from the calculated transition strength function.

\end{abstract}

\section{Introduction}
Spontaneous symmetry breaking is one of the important ingredients in the nuclear density functional theory (DFT).
Usually, self-consistent mean-field wave-function is obtained in the Hartree-Fock or Hartree-Fock-Bogoliubov (HFB) framework.
As it often happens, obtained mean-field wave-function can break some of the symmetries, which are, nevertheless, conserved by the
underlying Hamiltonian. 
For example, the variational space can be increased by allowing the nucleus to deform, leading to a lower ground state energy.
A systematic survey across the nuclear chart indeed indicates that most of the calculated nuclei turn out to be deformed~\cite{Erler2012}.
Effectively, this kind of symmetry breaking mechanism allows incorporating various correlations
on the mean-field wave function. 
The consequence of symmetry breaking via deformation is that the wave-function is no longer an eigenstate of the angular momentum operator.

In the linear response theory, that is, in the random-phase approximation (RPA) theory, a broken symmetry at the mean-field level
manifests as a zero-energy, symmetry restoring, Nambu-Goldstone (NG) mode. This mode does not represent an intrinsic excitation of 
the system, but, instead, it represents a special case of collective motion which needs to be treated separately. 
When computing certain kind of transition modes, the NG mode, associated with broken symmetry, appears as a solution of the RPA equations. 
For this reason, this mode is often called as a spurious mode.
Depending on the broken mean-field symmetry, various NG modes can appear \cite{ring80,Hinohara2015,Hinohara2016,Petrik2018}.
In the present work I will focus on the
translational NG mode, stemming from the broken translational invariance, and rotational NG mode, caused by deformed mean-field.

Traditionally the RPA and its superfluid variant, the quasiparticle-RPA (QRPA), have been formulated in the matrix form.
Unfortunately, a breaking of spherical symmetry leads to very large matrices and to keep calculation traceable,
various truncations are usually introduced. With iterative QRPA methods, the construction of a large matrix can be circumvented,
and the QRPA problem can be solved without resorting to any truncations. One such kind of method is the finite-amplitude-method 
(FAM)~\cite{Nakatsukasa2007}, which is also employed in the present work.

\section{Theoretical framework}
The purpose of the QRPA is to describe small amplitude oscillations around the HFB ground state. Consequently, the FAM-QRPA equations
can be derived by starting from the time-dependent HFB theory and taking a small amplitude limit.
In the FAM calculation, these oscillations are induced by applying an external polarizing field on the system,
which corresponds to an operator $F$. The FAM amplitudes are obtained as
\begin{equation}
X_{\mu\nu}(\omega)  =  -\frac{\delta H^{20}_{\mu\nu}(\omega)-F^{20}_{\mu\nu}}{E_{\mu}+E_{\nu}-\omega} \,, \quad
Y_{\mu\nu}(\omega)  =  -\frac{\delta H^{02}_{\mu\nu}(\omega)-F^{02}_{\mu\nu}}{E_{\mu}+E_{\nu}+\omega} \, , \label{eq:FAMXY}
\end{equation}
where $\delta H^{20}_{\mu\nu}$ and $\delta H^{02}_{\mu\nu}$ contain the induced fields coming from the underlying 
energy density functional (EDF), which encodes the nucleonic interactions. Since these fields depend on the FAM amplitudes, 
equations (\ref{eq:FAMXY}) needs to be solved iteratively for each value of $\omega$.

The strength function for the operator $F$ at the frequency $\omega$ is defined as
\begin{equation}
S(F;\omega) = \sum_{\mu\nu} \left[F_{\mu\nu}^{20\ast} X_{\mu\nu}(\omega)+F_{\mu\nu}^{02\ast}Y_{\mu\nu}(\omega)\right] \,,
\label{eq:Str}
\end{equation}
which can be then used to calculate transition strength function as
\begin{equation}
\frac{d B(F ;\omega)}{d \omega} = -\frac{1}{\pi} {\rm{Im}}\, S(F ;\omega)\,.
\end{equation}
To obtain a finite value for transition strength, the FAM equations are usually solved by introducing a small imaginary
component on the frequency $\omega \to \omega +i\gamma$.

The NG mode appears when the mean-field breaks some continuous symmetry. In this kind situation, there exists an operator $P$ 
which is a generator of the broken symmetry group.  For example, localization of the mean-field 
breaks the translational invariance, and, consequently, the HFB solution no longer commutes with the momentum operator.
As demonstrated in \cite{Hinohara2015}, in this kind of situation, the Thouless-Valatin (TV) inertia parameter associated with the operator 
$P$ can be obtained from the FAM-QRPA strength function at zero energy as $S(P;0)=-M_{\rm NG}$.
With NG mode, there exists also a conjugate operator $Q$ with a commutation relation $[Q,P]=i$.
Operators $P$ and $Q$ are needed together to form the RPA solution for the NG mode~\cite{ring80}.

The procedure to remove spurious translational mode in the FAM calculation was introduced in~\cite{Nakatsukasa2007}.
With this procedure, the physical amplitudes, $\{X_{\mu\nu}^{\rm phys},Y_{\mu\nu}^{\rm phys}\}$,
which are used to compute transition strength, are obtained from calculated amplitudes, 
$\{X_{\mu\nu}^{\rm calc},Y_{\mu\nu}^{\rm calc}\}$, with following correction
\begin{eqnarray}
X_{\mu\nu}^{\rm phys}(\omega) & = & X_{\mu\nu}^{\rm calc}(\omega) -\lambda_P P^{20} -\lambda_Q Q^{20} \,, \\
Y_{\mu\nu}^{\rm phys}(\omega) & = & Y_{\mu\nu}^{\rm calc}(\omega) -\lambda_P P^{02} -\lambda_Q Q^{02} \,.
\end{eqnarray}
The same method can be used also to remove other spurious modes.
Coefficients $\lambda_P$ and $\lambda_Q$ are obtained from the requirement that physical modes and spurious mode are orthogonal to each other.
This gives
\begin{eqnarray}
\lambda_P & = & N^{-1} \left( \langle Q^{20} \vert X^{\rm calc}(\omega) \rangle + \langle Q^{02} \vert Y^{\rm calc}(\omega) \rangle \right) \,,\\
\lambda_Q & = & -N^{-1} \left( \langle P^{20} \vert X^{\rm calc}(\omega) \rangle + \langle P^{02} \vert Y^{\rm calc}(\omega) \rangle \right) \,,
\end{eqnarray}
where the normalization constant is
\begin{equation}
N = \langle Q^{20},Q^{02}\vert P^{20},P^{02}\rangle \,.
\end{equation}
In principle, the normalization constant has a value of $N=i$. However, 
the use of a finite size basis may lead to a situation where $N$ deviates from this value. This is,
for example, the case with momentum and position operators in the harmonic oscillator basis.
Similar procedure to remove spurious mode was also introduced for iterative Arnoldi diagonalization method~\cite{Toivanen2010}.

There are cases when the analytic form of the operator $Q$ is not available. This is, for example, the case with rotational NG mode.
In this kind of situation, it was shown in~\cite{Hinohara2015}, that it
can be obtained from the response to the $P$ operator at zero energy as
\begin{equation}
Q_{\mu\nu}^{20} = -i\frac{X_{\mu\nu}(0)+Y_{\mu\nu}^{*}(0)}{2S(P;0)} \,. \label{eq:Qcon}
\end{equation}

\section{Results}
To demonstrate how TV inertia parameter can be obtained from FAM calculation, and how spurious mode can be removed
from transition strength function, I have taken nucleus $^{168}$Er as an example. 
The HFB solution was obtained with the computer code HFBTHO~\cite{Stoitsov2013}, by using a basis consisting of 
20 major oscillator shells and the SkM* Skyrme EDF~\cite{SKMS} with mixed pairing description at the pp-channel. 
Since this nucleus has a stable prolate deformed HFB ground state, with deformation parameter $\beta=0.32$, 
the $K^{\pi}=1^{+}$ modes will have an admixture of spurious rotational mode. 
In addition, the localization of the mean-field leads to the appearance of spurious $K^{\pi}=0^{-},1^{-}$ translational modes.
The subsequent FAM calculations were handled with the FAM module developed in \cite{Kortelainen2015}. 
As noted earlier, the FAM calculations can be performed without any truncations.

\begin{figure}[h]
\begin{center}
\includegraphics[width=0.99\textwidth]{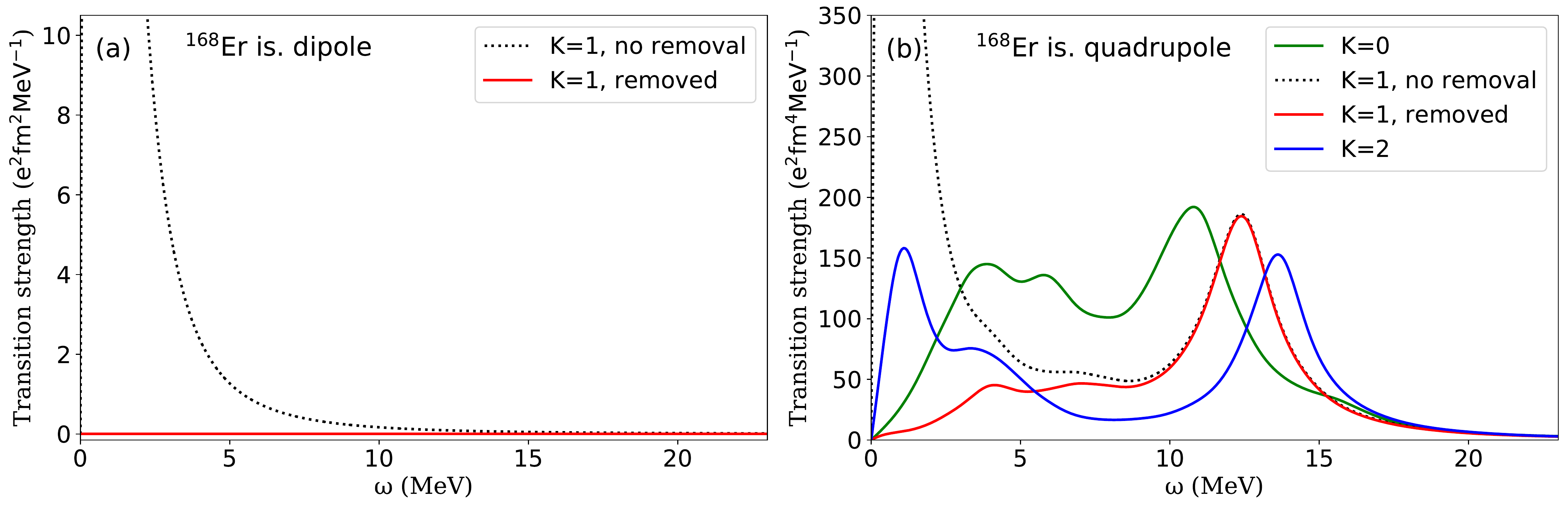}
\caption{\label{fig:isdq} 
Calculated transition strength in $^{168}$Er, with SkM* Skyrme EDF,
(a) for isoscalar dipole $K=1$ mode with and without removal of the spurious translational mode, and
(b) for isoscalar quadrupole for $K=$ 0, 1, and $2$ modes. 
For $K=1$ case, the transition strength is shown with and without removal of the spurious rotational mode.}
\end{center}
\end{figure}

With use of $K=1$ momentum operator at frequency $\omega=0$, the obtained TV inertia mass parameter was $M_{\rm NG}=4.074379\,{\rm MeV^{-1}\,fm^{-2}}$.
This corresponds to mass number $A_{\rm NG}\approx 167.951$, deviating only $0.03$\% from the $A=168$.
Figure~\ref{fig:isdq}, panel (a), shows calculated isoscalar, $K=1$, dipole transition strength in $^{168}$Er. In this case,
the only nonzero mode is the spurious mode, depicted as a dotted line in the panel. Once the spurious mode is removed,
the remaining transition strength vanishes. The same result is obtained either by taking $Q$ as a position operator or 
by constructing $Q$ as in equation (\ref{eq:Qcon}). The inertia parameter and removal of spurious mode can be handled 
for $K=0$ translational mode in the same manner. After removal of the spurious mode, the $K=0$ isoscalar dipole transition 
strength vanishes.

By applying $J_{\rm y}$ operator on the FAM calculation with $\omega=0$, the rotational NG can be addressed. As a result, I have
obtained rotational moment of inertia $M_{\rm NG}=34.917269\,{\rm MeV^{-1}}$ for $^{168}$Er. In a simple 
rotational picture, this corresponds to $2^+$ state energy of $E(2^+)=86\,{\rm keV}$ for the ground state rotational band.
Experimentally, the energy of the $2^+$ state, belonging to the ground state rotational band, is $79.8\,{\rm keV}$~\cite{IAEA}.
As was show in~\cite{Petrik2018}, the obtained TV inertia coincides with the one obtained from self-consistent
cranking calculation at the limit of zero cranking frequency.

With the rotational case, the analytic form of the operator $Q$ is not known, and the method in equation (\ref{eq:Qcon})  should
be used to obtain the conjugate operator, to remove the spurious rotational mode.
This is demonstrated in figure~\ref{fig:isdq}, panel (b), showing the calculated isoscalar quadrupole $K=1$ transition 
strength before and after removal of the spurious mode. As can be seen, after the removal procedure, the low-lying 
spurious strength has been removed, whereas the impact at higher excitation energy is minuscule.
Panel (b) also shows calculated transition strength for isoscalar $K=0$ and $K=2$ quadrupole modes.
Due to deformation effects, the position of the giant resonance splits and becomes different for each $K$-mode.

\section{Conclusions and discussion}
The FAM formalism allows computing TV inertia parameter, which can be utilized, e.g., as a
microscopical input for various collective Hamiltonian models~\cite{Hinohara2010,Washiyama2017}. In addition to this, I have demonstrated that
translational and rotational spurious mode can be removed within the current FAM formalism.
This allows computation of the transition strength function, which is free from spurious mode, for any electromagnetic operator.

\section*{Acknowledgements}
I thank Nobuo Hinohara for useful discussions.
This work was supported by the Academy of Finland under the Academy project no. 318043.
I acknowledge CSC-IT Center for Science Ltd., Finland, for the allocation of computational resources.

\section*{References}
\providecommand{\newblock}{}

\end{document}